# Hypothesis of collective state in biomolecules and discussion on the possibility of its experimental verification

Anatoly M. Smolovich

*Kotel'nikov Institute of Radio Engineering and Electronics (IRE) of the Russian Academy of Sciences, Moscow, Russia*

*e-mail: asmolovich@petersmol.ru*

**Abstract:** The analogy between the phenomenon of live and collective states in Condensed Matter was supposed by several well-known physicists. The possibility of experimental verification of this hypothesis is discussed.

In the Forward to the first edition of his monograph *Superfluids* [1] F. London supposed the possibility of some quantum behavior akin to superfluidity that might play a role in biological processes. He said that certain actions between macromolecules in biochemistry could be understood unless they could be conceived by some quantum mechanism involving the system as whole. It is conceivable that in some biological processes the concept of fluid state of entropy zero could play a decisive role, for it combines the characteristic stability of quantum states with the possibility of motion (i.e. of matter transfer and change of shape) without necessarily implying dissipation processes. In [2] W.A. Little paid attention on this remark of Fritz London and noted that if this should be the case, an entirely new and important consideration would be added to the problem of understanding living systems. Then, W.A. Little suggested that superconductivity near room temperature might be achieved in an organic polymer, which was loosely related to the structure of DNA. In [3] H. Frohlich conjectured that some collective quantum effects can take place in living systems. My hypothesis is that some special type of macroscopic quantum phenomena (also called collective states) is the basic property the life phenomenon itself. The question is how to verify this hypothesis in experiment. This is the subject of a following discussion.

I suppose that here the presence of energy gap can be used as an indicator of some collective state presence. For superconductors the energy gap is a region of suppressed density of states around the Fermi energy. The superconducting energy gap gives stability to superconductivity because it can be considered as a portion of energy that should be applied to destroy superconductivity. The size of the energy gap indicates the energy gain for two electrons upon formation of a Cooper pair. BCS theory predicts that the size $\Delta$ of the superconducting energy gap for conventional superconductors $\Delta(T=0) = 1.76\ k_B T_c$, where $T_c$ is the temperature superconducting transition and $k_B$ is Boltzmann constant [4]. The energy gap width of high temperature superconductors [5] and charge density wave (CWD) [6] differs but also have an order about $k_B T_a$, where $T_a$ is the temperature of superconducting or Peierls transition, respectively.

My hypothesis is that the life phenomenon is also characterized by presence of some correspondence energy gap. We don't know is it true or not but we can try to search for this energy gap in an experimental investigation. I suppose that the width of this energy gap has an order of $k_B T_V$, where $T_V$ is some temperature typical for living organisms, for example 300K. This gives the estimation for energy gap about 0.026 eV.

Some simple organisms should be chosen for this research. Indeed, even simple living organisms like unicellulate, prokaryotes or viruses are significantly more complicated than the majority of physical research subjects. All organisms contain DNA or RNA, which store the biological information. The simplest organisms are viruses, which consist of nucleic acid (DNA or RNA) surrounded by a protective coat of protein called a capsid. However, opinions differ on whether viruses are a form of life or organic structures that interact with living organisms. They have been described as "organisms at the edge of life." In [7] viruses are considered to act an important part in the evaluation process. The viruses don't demonstrate the life characters as metabolism, replications, etc. when they are out of the host-cell. This allows us to use the phage's state inside and outside the host-cell as markers of alive (V, vita) and dead (M, mort) states [8]. Besides, viruses can be simply inactivated, for example by UV illumination or by heating. The inactivated virus doesn't demonstrate the life characters inside the host-cell. We don't know exactly where the border between V and M virus states is situated. We can't a priory exclude that this border is located between the normal and inactivated states of virus. The electronic energy levels of these states should be compared. The electronic density of states (DOS) of the viruses inside and outside of the host-cell should be measured.

Now, let's discuss what experimental methods can be used for identification of the energy gap. Tunnelling spectroscopy is the appropriate technique for measuring the DOS including identification of the superconducting and CDW energy gaps [9, 10]. Scanning tunnelling spectroscopy (STS) is very useful for various molecular objects and nanoparticles deposited on substrates. Also, there were attempts of single DNA molecules STS investigation across the helices [11-16]. However, its clear interpretation was inhibited by technical hurdles including a great data spread. There is another problem. As I have mentioned, it is desirable to compare the results of spectroscopic measurements of the phages outside and inside of the host-cells. However, the possibility of STS investigation of phages inside the host-cell casts doubt.

The other method I would like to mention is the traditional optical spectroscopy in different wavelength region of the electromagnetic radiation spectrum. The first spectroscopic measurement of the superconducting energy gap was made with microwaves [17] and far infrared techniques [18]. This spectroscopy also was used for investigations of biomolecules [19-22]. Note, that interpretation of the biomolecules optical spectrum is very difficult task due to the complicated structure of these molecules. The additional problem is that the size of light spot significantly exceeds the thickness of DNA molecule. This leads to necessity of desired signal measuring against the strong background noise.

I see the way for overcoming these difficulties in using of both STS and optical spectroscopy methods for the same DNA molecules. Comparison of the results obtained by these methods should help in their interpretation. The other point is comparison of the spectroscopy results for V and M virus states. The zone of spectra where some difference between V and M virus states will be found should be the subject of farther detailed investigations.

For decision of research methodology let's pay attention on publications [23-25] where the virus UV inactivation was investigated. The bacteriophages Lambda (λ) and phi X 174

(ΦX174) were used in the experiments. The virus inactivation after UV irradiation was checked by phages seeding on the corresponding strains of Escherichia coli (host-cell).

Now, I would like to propose the following research methodology. Both phages and corresponding host-cells should be investigated to find the difference in the DOS between V and M states of phages. The border between V and M states of phages can be situated between the normal and inactivated phages or between the normal phage inside and outside the host-cell. We should check both options. Firstly, the spectroscopy of host-cells should be performed before their infecting by phages. The spectroscopy of phages in normal state should be performed both outside and inside the host-cell. The spectroscopy of phages outside the host-cell should be done by two different methods: STS and optical spectroscopy. In [26] the discrimination of bacteria and bacteriophages by Raman spectroscopy and surface-enhanced Raman spectroscopy was demonstrated. In [27] detection and discrimination of viruses was fulfilled by the use of Fourier transform infrared spectroscopy. Also see the subsequent publications [28-33]. The spectroscopy of phages inside the host-cell should be done only by optical spectroscopy. The same measurement of the UV inactivated phages should be done. The DOS in V and M states of the phages should differ even in case if the hypothesis of energy gap existence is wrong. The difference between the DOS of V and M phages states can be found by comparison of the measured data. The specific energy gap of living state can be possibly identified by this way.

In summary, the possibility of experimental verification of hypothesis of some macroscopic quantum state in living organisms was discussed. I thank A.A. Sinchenko and S.V. Chekalin for useful consulting and discussions.